\documentclass{svjour3}

\usepackage[utf8]{inputenc} 
\usepackage{graphicx}
\usepackage{amsmath}
\usepackage{amsfonts}
\usepackage{amssymb}
\usepackage{mathtools}
\usepackage{float}
\usepackage{tabularx, booktabs}
\usepackage{color}
\usepackage{braket}
\usepackage{bm}
\usepackage{comment}

\usepackage{hyperref}

\hypersetup{
	colorlinks=true,
	linkcolor=blue,
	citecolor=blue,
	filecolor=magenta,
	urlcolor=cyan,
	breaklinks=true
}
\begin{document}

\title{Absence of binding of heavy fermions by two light atoms in two dimensions}

\author{J. Givois         \and
        D. S. Petrov
}

\institute{J. Givois \and D. S. Petrov \at
              Universit\'e Paris-Saclay, CNRS, LPTMS, 91405 Orsay, France \\
}

\maketitle

%\date{\today}

\begin{abstract}
By developing the mean-field theory valid for large $N$, we investigate the problem of two light fermions interacting via a zero-range potential with $N$ heavy fermions in two dimensions. We obtain numerical evidence that this system is never fully bound. It always splits into droplets containing a single light atom. This is in contrast to the one-dimensional case where any number of heavy and light fermions can be bound together.

\end{abstract}

\section{Introduction}\label{Sec:Intro}

The question how many identical heavy fermions can be bound together by a single light atom has been extensively discussed in recent years. This $N+1$-body problem is characterized by a competition between the Pauli repulsion (or Fermi pressure) of the heavy component and the effective attraction due to the fact that the light atom gains in energy if the attractive scatterers are close to one another. This problem is controlled by the number of the heavy atoms $N$, the heavy-light mass ratio $M/m$, and the space dimension. The binding gets systematically easier for smaller $N$, for larger $M/m$, and in lower dimensions. The heavy-light zero-range attraction is parametrized by a positive scattering length $a$, which defines the dimer size and sets the length scale of the problem.

Exact few-body calculations of the $N+1$-body problem have been carried out up to the $4+1$ pentamer in two and three dimensions and up to the $5+1$ hexamer in one dimension. In three dimensions, the $2+1$ trimer binds at $M/m = 8.2$~\cite{KartavtsevMalykh}, the $3+1$ tetramer at $M/m = 8.9$~\cite{Blume,bazak2017}, and the $4+1$ pentamer at $M/m = 9.7$~\cite{bazak2017}. A peculiarity of three-dimensional $N+1$-body systems is that they become Efimovian for mass ratios above $N$-dependent thresholds around $M/m\approx 13$~\cite{Efimov,Castin,bazak2017}. In two dimensions the $2+1$ trimer emerges for $M/m>3.33$~\cite{PricoupenkoPedri}, the $3+1$ tetramer for $M/m>3.38$~\cite{liu2022}, and the $4+1$ pentamer for $M/m>5.14$~\cite{liu2022}. In one dimension the $2+1$ trimer exists for any $M/m>1$~\cite{Kartavtsev} and the mass-ratio thresholds of larger $N+1$ clusters are $(M/m)_{3+1} = 1.76$, $(M/m)_{4+1} = 4.2$, and $(M/m)_{5+1} = 12.0$~\cite{tononi2022}.

To describe $N+1$ clusters for large $N$ in one and two dimensions we have developed the mean-field (MF) theory combined with the Thomas-Fermi (TF) approximation for the heavy fermions~\cite{tononi2022,givois2023,givois2023_2}. By properly rescaling the corresponding energy functional one observes that the MF solution essentially depends on the parameters $N$ and $M/m$ only through the combinations $\alpha_{\rm 1D}=(\pi^2/3)N^3/(M/m)$ and $\alpha_{\rm 2D}=4\pi N^2/(M/m)$ in one and two dimensions, respectively. The $N+1$-cluster is bound, respectively, for $\alpha_{\rm 1D}<12$ and $\alpha_{\rm 2D}<11.7$. Equivalently, one can say that the $N+1$ cluster is bound and can still accept more heavy atoms if $N<N_c$, where $N_c=0.965 (M/m)^{1/2}$ in two dimensions and $N_c=1.539 (M/m)^{1/3}$ in one dimension. At $N=N_c$ the cluster is saturated and cannot bind more heavy atoms. %We call $N/N_c$ the saturation ratio of a $N+1$ cluster. 

Applying this MF theory in one dimension we have also considered the case of more than one light fermion and found that this system can be self bound~\cite{givois2023}. This result is not obvious since the subsystem of the light fermions is characterized by its own Fermi pressure, which is strong ($\propto 1/m$) and goes against binding. We do find that two $N+1$ droplets always repel each other at large distances, but at smaller separations they may or may not attract depending on their saturation ratio $N/N_c$. When the number of light atoms $N_l\geq 2$ the system can bind as a chain of length $N_l$ of $N+1$ droplets with $N$ equal to the number of heavy fermions per light atom.

In this paper we report on our numerical analysis of the $N_l=2$ case in two dimensions generalizing the technique of Ref.~\cite{givois2023_2} to two light fermions. Our main finding is that the system is never fully bound and results of our numerics are consistent with the following scenario. As we add more heavy atoms at a fixed (large) $M/m$ we pass the following three stages. First, for $N<N_c$ we are dealing with a single $N+1$ droplet plus a free light atom. Then, for $N_c<N<2 N_c$ the first droplet is saturated and the second droplet starts being formed around the second light atom. This second droplet contains $N-N_c$ heavy fermions and it is much larger in size than the saturated one. Finally, for $N>2N_c$ the state of the system is two unbound identical saturated droplets plus a free gas of $N-2N_c$ heavy fermions. We can thus conclude that for sufficiently large $N$ and $M/m$ (which are the validity requirements of the MF theory) the $N+2$ cluster cannot be bound by purely zero-range interspecies forces. %We attribute this phenomenon to a more decisive role of the centrifugal barrier for identical fermions in two dimensions than in in one dimension.
As a reservation, we cannot exclude that such clusters with finite $N$ may be bound due to beyond-MF effects. At the moment we also cannot rule out binding of $N+N_l$ clusters with $N_l>2$. 

The paper is organized as follows. In Sec.~\ref{Sec:MF} we derive the MF equations for the $N+N_l$ system valid for $N\gg 1$ and $N_l\sim 1$. An important feature of this two-dimensional MF theory is that self-bound stationary solutions always correspond to vanishing energy. The MF theory is thus not enough for determining binding energies of various cluster configurations and comparing them to each other. Therefore, for this analysis it is necessary to take into account beyond-MF effects. In Sec.~\ref{Sec:BMF} we explain how to do this and we recall the main results obtained for $N_l=1$. In Sec.~\ref{Sec:Nplus2} we discuss the $N+2$ case and present our numerical results. In Sec.~\ref{Sec:Conc} we conclude.

\section{Mean-field approach}\label{Sec:MF}

We start with the general Hamiltonian of our mass-imbalanced Fermi-Fermi mixture
\begin{equation}\label{eq:Main_Hamiltonian}
\hat{H} = \int \bigg( -\frac{\hat{\phi}^{\dagger}_{{\bf r}} \nabla_{{\bf r}}^2 \hat{\phi}_{{\bf r}} }{2m}-\frac{\hat{\Psi}^{\dagger}_{{\bf r}} \nabla_{{\bf r}}^2 \hat{\Psi}_{{\bf r}} }{2M}  + g \hat{\Psi}^{\dagger}_{{\bf r}} \hat{\phi}^{\dagger}_{{\bf r}}  \hat{\Psi}_{{\bf r}} \hat{\phi}_{{\bf r}} \bigg) d^2 r.
\end{equation}
Here $\hat{\phi}_{\bf r}^{\dagger}$ and $\hat{\Psi}_{\bf r}^{\dagger}$ are the creation operators of light and heavy fermions and we set $\hbar=1$. The zero-range interactions between heavy and light atoms are modeled with the help of the coupling constant 
\begin{equation}\label{g}
g=2\pi/[m_r\ln(2m_r|E_{1+1}|/\kappa^2)]<0
\end{equation} 
and an ultraviolet cut-off momentum $\kappa$ much larger than any other momentum scale in the problem. In Eq.~(\ref{g}) we also introduce the reduced mass $m_r=mM/(m+M)$ and the dimer energy $E_{1+1}$ related to the heavy-light scattering length by $E_{1+1}=-2e^{-2\gamma_E}/(m_r a^2)$, where $\gamma_E$ is the Euler constant. 

We write the MF Hartree-Fock energy functional for the system containing $N$ heavy and $N_l$ light atoms as
\begin{equation}\label{MFenergyHF}
E_{\rm HF}=\int \left[\sum_{i=1}^{N_l}|\nabla \phi_i({\bf r})|^2/(2m)+\sum_{j=1}^{N}|\nabla \Psi_j({\bf r})|^2/(2M) +g n_h({\bf r})n_l({\bf r})\right] d^2 r,
\end{equation}
where $\phi_i({\bf r})$ and $\Psi_j({\bf r})$ are the Hartee-Fock orbitals for the light and heavy atoms, respectively, and we introduce the densities $n_h({\bf r})=\sum_{j=1}^{N}|\Psi_j({\bf r})|^2$ and $n_l({\bf r})=\sum_{i=1}^{N_l}|\phi_i({\bf r})|^2$. Equation~(\ref{MFenergyHF}) is the variational energy corresponding to the product of Slater determinants built on orthonormal sets $\phi_i({\bf r})$ and $\psi_j({\bf r})$. This is a good approximation to the true energy for weak interactions, i.e., for $m_r |g|\ll 1$. The minimization of Eq.~(\ref{MFenergyHF}) with normalization constraints for the orbitals gives the equations
\begin{equation}\label{Schrphi}
-\nabla^2 \phi_i({\bf r})/(2m) + g n_h({\bf r}) \phi_i({\bf r}) = \epsilon_i\phi_i({\bf r})
\end{equation}
and
\begin{equation}\label{HForbitals}
-\nabla^2 \Psi_j({\bf r})/(2M)+gn_l({\bf r}) \Psi_j({\bf r})=\omega_j \Psi_j({\bf r}),
\end{equation}
where $\epsilon_i$ and $\omega_j$ are Lagrange multipliers. The interpretation of the Hatree-Fock equations (\ref{Schrphi}) and (\ref{HForbitals}) is that each component is an ideal Fermi gas placed in the ``external'' trapping potential equal to $g$ multiplied by the density of the other component. These equations can be efficiently solved by iterations.

We will be interested in the case $N/N_l\gg 1$ and we will use the local-density Thomas-Fermi approximation for the heavy fermions valid when $n_h$ changes slowly on the mean interparticle distance scale. In this approximation the kinetic energy density of the heavy atoms $1/(2M)\sum_{j=1}^{N}|\nabla \Psi_j({\bf r})|^2$ is replaced by the ground state energy density $\pi n_h^2({\bf r})/M$ of a uniform Fermi sea of density $n_h({\bf r})$. The energy functional of the system then reads
\begin{equation}\label{MFenergy}
E=\frac{1}{2m}\int \left\{\frac{\alpha}{2} n^2({\bf r})+\sum_{i=1}^{N_l} \left[|\nabla \phi_i({\bf r})|^2 +\gamma n({\bf r})|\phi_i({\bf r})|^2\right]\right\} d^2 r,
\end{equation}
where we rescale the density of the heavy atoms $n_h({\bf r})=N n({\bf r})$ and introduce the dimensionless parameters $\alpha=4\pi m N^2/M$ and $\gamma=2mgN<0$. % that we have already used in Ref.~\cite{givois2023_2}. 
From Eq.~(\ref{MFenergy}) one can see that apart from the overall prefactor $1/(2m)$ the energy functional is controlled by the mass ratio and $N$ only through the combination $\alpha$. 

The minimization of Eq.~(\ref{MFenergy}) with respect to $\phi_i$ leads to an equation equivalent to Eq.~(\ref{Schrphi}). In new notations it explicitly reads
\begin{equation}\label{SchrphiNN}
-\nabla^2 \phi_i({\bf r}) + \gamma n({\bf r}) \phi_i({\bf r}) = 2m\epsilon_i\phi_i({\bf r})
\end{equation}
Minimizing Eq.~(\ref{MFenergy}) with respect to $n({\bf r})$ leads to
\begin{equation}\label{Dens}
	n({\bf r}) = -\frac{\gamma}{\alpha} \theta[\sum_{i=1}^{N_l}|\phi_i({\bf r})|^2 +2mN \mu/\gamma],
\end{equation}
where $\theta(x)=(x+|x|)/2$ and $\mu$ is the chemical potential of the heavy atoms, which is the Lagrange multiplier corresponding to the normalization condition $\int n({\bf r}) d^2 r = 1$.

The sets Eqs.~(\ref{Schrphi})-(\ref{HForbitals}) and Eqs.~(\ref{SchrphiNN})-(\ref{Dens}) share the following scaling property. Let us consider, for concreteness, the Thomas-Fermi case Eqs.~(\ref{SchrphiNN})-(\ref{Dens}) and assume that $\phi_i({\bf r})$, $n({\bf r})$, $\epsilon_i$, and $\mu$ solve these equations for a certain $\gamma$. Then, for any $\lambda$ the combination $\lambda \phi_i(\lambda {\bf r})$, $\lambda^2 n(\lambda {\bf r})$, $\lambda^2\epsilon_i$, and $\lambda^2\mu$ is also a correctly normalized solution corresponding to the same $\gamma$. This two-dimensional MF theory is peculiar in the sense that, independently of $\lambda$, all the self-similar solutions correspond to vanishing total energy Eq.~(\ref{MFenergy}). This can be shown directly by manipulating with Eqs.~(\ref{SchrphiNN}) and (\ref{Dens}), with their derivatives with respect to $\lambda$, and by partial integration~\cite{givois2023_2}. One can also understand this qualitatively by noting that the total energy $E$ also scales proportionally to $\lambda^2$ and, if $E$ were not zero, the system would prefer to expand or collapse. The relation between the stationarity condition and vanishing of $E$ has been established in Ref.~\cite{Vlasov} for the problem of light beam propagating through a nonlinear medium. Another peculiarity of Eqs.~(\ref{SchrphiNN}) and (\ref{Dens}) is that once $\alpha$ is fixed, the solution exists only for a certain $\gamma=\gamma_c(\alpha)$. More precisely, at this critical $\gamma$ we get the whole family parametrized by $\lambda$. 

One can easily show that the Hartree-Fock Eqs.~(\ref{Schrphi})-(\ref{HForbitals}) possess the same scaling property. Indeed, instead of $n({\bf r})$ and $\mu$ we deal with $\Psi_j({\bf r})$ and $\omega_j$ which rescale as $\Psi_j({\bf r})\rightarrow \lambda \Psi_j(\lambda {\bf r})$ and $\omega_j\rightarrow \lambda^2\omega_j$. However, the critical $\gamma_c^{\rm TF}$ and $\gamma_c^{\rm HF}$ obtained, respectively, with the Hartree-Fock and the Thomas-Fermi methods are different but should approach each other in the asymptotic limit of large $N$. Indeed, in Ref.~\cite{givois2023_2} we numerically show that for the $N+1$ cluster problem at fixed $\alpha$ the difference $\gamma_c^{\rm HF}-\gamma_c^{\rm TF}\sim 1/N$ at large $N$. We point out that in the Thomas-Fermi approximation $\gamma_c^{\rm TF}$ depends only on $\alpha$, whereas in the Hatree-Fock treatment $\gamma_c^{\rm HF}$ depends separately on $N$ and $M/m$.

The validity condition of the MF assumption is the regime of weak interactions, i.e.,  $m_r|g|\ll 1$. In addition, the Thomas-Fermi assumption requires $N\gg 1$. Both these two requirements are satisfied in the thermodynamic limit defined by the limiting procedure implying $N\rightarrow \infty$, $M/m\rightarrow \infty$, and $g\rightarrow 0$ such that $\alpha$ and $\gamma$ are constant. 

\section{Beyond-mean-field term and determination of size and energy}
\label{Sec:BMF}

The MF treatment does not predict the size of the bound state nor the binding energy. These quantities have to be determined by taking into account the beyond-MF correction to the energy, which is a local term calculated in the second-order perturbation theory~\cite{givois2023_2}. It is local since the corresponding integral is dominated by the contribution of virtual excitations with momenta larger than the (local) Fermi momentum of the heavy atoms. Adding this term to the MF Eq.~(\ref{MFenergyHF}) or to Eq.~(\ref{MFenergy}) is equivalent to replacing the coupling constant $g$ by its renormalized value, which depends on the local density of the heavy atoms
\begin{equation}\label{grenorm}
g\rightarrow g-m_r g^2\frac{\ln[e^{1/2}\kappa/p_F({\bf r})]}{\pi},
\end{equation}
where $p_F({\bf r})=\sqrt{4\pi n_h({\bf r})}$.

From Eq.~(\ref{grenorm}) we see that the beyond-MF correction is weaker than the MF interaction term by the factor $m_r |g|$, which is indeed small in the thermodynamic limit defined in the end of Sec.~\ref{Sec:MF}. Note that $m_r |g|\sim 1/N \ll 1$ since $\alpha$ and $\gamma$ are kept finite. The hierarchy of energy scales in the thermodynamic limit is thus as follows. The MF interaction and the kinetic energy of a bound state cancel each other for $\gamma=\gamma_c$, but each of these terms is of order $N^0$ and it is this energy scale that determines the shapes of $\phi_i$ and $n$. The beyond-MF term is proportional to $1/N$. It is too weak to induce significant changes in $\phi_i$ and $n$. However, it breaks the degeneracy associated with the scaling parameter $\lambda$ and determines the energy of the cluster. The procedure for finding the optimal $\lambda$ and the corresponding $E$ is as follows. 

We solve the MF Eqs.~(\ref{SchrphiNN}) and (\ref{Dens}) selecting one member of the family of self-similar solutions. The concrete choice is not important, but we fix the length scale by setting $2m\epsilon_1=-1$, where $\epsilon_1$ is the energy of the lowest orbital for the light atoms. This constraint is practical as the localization length of the light fermion and the cluster size are then of order one. As a result of iteratively solving Eqs.~(\ref{SchrphiNN}) and (\ref{Dens}) we have the critical $\gamma_c$ (which does not depend on the length scale) and the dimensionless fields $\phi_i({\bf r})$ and $n({\bf r})$ as functions of the dimensionless ${\bf r}$. It is convenient to think of $\lambda$ as dimensional momentum or inverse size as the dimensional solutions with size $1/\lambda$ are characterized by the fields $\lambda \phi_i(\lambda {\bf r})$ and $\lambda^2 n(\lambda {\bf r})$. Note that $\gamma_c$ is not an external parameter, but a quantity determined by solving the MF equations. By contrast, $\gamma$ is an external parameter which we control. To have a stationary MF solution we should have $\gamma\approx \gamma_c$ with the MF accuracy, i.e.,
\begin{equation}\label{gammagammac}
|\gamma-\gamma_c|/|\gamma_c|\sim 1/N.
\end{equation}
Equation~(\ref{gammagammac}) means that we can tune $\gamma$, but keep its deviation form $\gamma_c$ such that the MF energy (proportional to $\gamma-\gamma_c$) remains smaller or comparable to the beyond-MF energy scale $\sim 1/N\ll 1$. Substituting $\lambda \phi_i(\lambda {\bf r})$, $\lambda^2 n(\lambda {\bf r})$, and Eq.~(\ref{grenorm}) into Eq.~(\ref{MFenergy}) we obtain the energy of the cluster up to beyond-MF terms as a function of $\lambda$ in the form
\begin{equation}\label{MFplusBMF}
E=\frac{I_1\gamma_c^2\lambda^2}{8\pi Nm}\left(4\pi N\frac{\gamma-\gamma_c}{\gamma_c^2}+\frac{I_2}{I_1}-\ln\frac{e\kappa^2}{4\pi N\lambda^2}\right),
\end{equation}
where $I_1=\int n(r)n_l(r)d^2 r$ and $I_2=\int n(r)n_l(r)\ln n(r) d^2 r$. From Eq.~(\ref{gammagammac}) it follows that with the beyond-MF accuracy $(\gamma-\gamma_c)/\gamma_c^2\approx 1/\gamma_c-1/\gamma$ and $4\pi N/\gamma\approx\ln[4e^{-2\gamma_E}/(a\kappa)^2]$. Then, minimization of Eq.~(\ref{MFplusBMF}) gives 
\begin{equation}\label{Rmin}
\lambda_{\rm min}^2=(\pi N a^2)^{-1} e^{-4\pi N/\gamma_c-I_2/I_1 -2\gamma_E }
\end{equation}
and
\begin{equation}\label{Enmin}
E=-\frac{I_1\gamma_c^2\lambda_{\rm min}^2}{8\pi Nm}.
\end{equation}

From Eqs.~(\ref{Rmin}) and (\ref{Enmin}) we see that the leading-order dependence of the cluster energy and size ($\propto 1/\lambda$) on $N$ is exponential. In addition, these equations also predict the algebraic terms (preexponential factors). In Ref.~\cite{givois2023_2} we have shown that to keep up with the preexponential accuracy one has to determine $\gamma_c$ by the Hartree-Fock method, which treats the kinetic energy of the heavy fermions more precisely. We have found numerically that $\gamma_c^{\rm TF}-\gamma_c^{\rm HF}\propto 1/N$ for large $N$. This difference thus influences the preexponential factor. 

Here we are interested in the leading-order exponential behavior and do not go beyond the asymptotic formula
\begin{equation}\label{Enminlead}
E\approx-m^{-1}a^{-2}e^{-4\pi N/\gamma_c},
\end{equation}
where it is sufficient to set $\gamma_c=\gamma_c^{\rm TF}$. Curiously, Eq.~(\ref{Enminlead}) is obtained from the beyond-MF theory, but its main ingredient $\gamma_c$ results from solving the MF Eqs.~(\ref{SchrphiNN}) and (\ref{Dens}). %a tool for analyzing which cluster configuration is energetically favorable just by comparing the corresponding $\gamma_c$. Indeed, for fixed $N$ and $\alpha$ the solution with larger $\gamma_c$ ($\gamma_c<0$) is more stable.

Before proceeding to the discussion of the $N+2$ case let us briefly recall some results of Ref.~\cite{givois2023_2} obtained in the case of a single light atom. The $N+1$ cluster exists for any $\alpha$ in the range from 0 to 11.7. The limit of small $\alpha$ corresponds to very high mass ratio where the heavy fermions are strongly localized, the light atom being in the halo state. Using the fact that the light atom wave function changes little inside the heavy cloud one can solve Eqs.~(\ref{SchrphiNN}) and (\ref{Dens}) analytically arriving at the asymptotic result $\gamma_c\approx 4\pi/\ln \alpha$ for small $\alpha$. 

Increasing $\alpha$ is equivalent to adding heavy atoms. This is possible till $\alpha=11.7$ where the chemical potential of the heavy atoms reaches zero and one can no longer add heavy particles. We call it the saturated $N_c+1$ droplet where $N_c=0.965(M/m)^{1/2}$. At this point $\mu=0$ and we have $n(r)= -\gamma|\phi(r)|^2/\alpha$. Then, Eq.~(\ref{SchrphiNN}) reduces to a nonlinear Schr\"odinger equation for $\phi$ with attractive cubic nonlinearity. The solution of this equation is known as the Townes soliton~\cite{Townes1964}. It corresponds to $\gamma_c=-\alpha=-11.7$.

The chemical potential of the heavy atoms can be calculated by taking the partial derivative of Eq.~(\ref{Enminlead}) with respect to $N$ at fixed $M/m$. Explicitly,
\begin{equation}\label{mu}
\mu=\partial{E_{N+1}}/\partial N\approx 8\pi [\alpha\gamma'_c/\gamma_c^2-1/(2\gamma_c)] E_{N+1},
\end{equation}
where $\gamma'_c=d\gamma_c/d\alpha$. One can see that the chemical potential is negative when $\gamma'_c/\gamma_c<1/(2\alpha)$ and the saturated droplet is characterized by $\gamma'_c=\gamma_c/(2\alpha)=-1/2$. These points are important for understanding what happens in the $N+2$ case.

\section{Case $N+2$}
\label{Sec:Nplus2}

Discussing the case of two light atoms we first mention that in the one-dimensional case the configuration of well-separated $N_1+1$ and $N_2+1$ clusters (with fixed $N_1+N_2$) reaches energy minimum when it is balanced, i.e., when $N_1=N_2$~\cite{givois2023}. By contrast, we find that in two dimensions one cluster always prefers to get to saturation by taking away all available heavy atoms from the other cluster. This happens because of the exponential scaling of the energy with $N$. The energy of the less populated droplet is exponentially negligible (in the thermodynamic limit) and it simply acts as a source of heavy fermions for the more populated cluster which tries to reach saturation. The only stable balanced configuration of two clusters is when there is enough heavy atoms to saturate both of them. We also investigate the stability of the balanced configuration by studying the behavior of $E_{(N+\delta N)+1}+E_{(N-\delta N)+1}$ for small $\delta N$. In this case we write
\begin{equation}\label{SumEn}
E_{(N+\delta N)+1}+E_{(N-\delta N)+1}\approx 2E_{N+1}+\delta N^2\partial^2 E_{N+1}/\partial N^2,
\end{equation}
where
\begin{equation}\label{SecDer}
\partial^2 E_{N+1}/\partial N^2= 64 \pi^2[ \alpha \gamma_c'/\gamma_c^2-1/(2\gamma_c)]^2 E_{N+1}<0.
\end{equation}
We indeed find that the balanced configuration is always unstable except when the term in the square brackets in Eq.~(\ref{SecDer}) is zero. As one can see from Eq.~(\ref{mu}) this is the saturation point where $\mu$ vanishes.

Although the above analysis assumes that the clusters do not overlap (but can exchange atoms), this simple picture well describes what we observe by numerically solving Eqs.~(\ref{SchrphiNN}) and (\ref{Dens}) for two light atoms. The main result of these calculations is that the system is never fully bound. A practical complication associated to this fact is that the system has to be confined. Therefore, in addition to the constraint $2m\epsilon_1=-1$, which sets the size of one of the clusters, we also require that all the fields vanish at the edge of a box of a variable size $L\times L$. Monitoring what happens with the solution when we increase $L$ we clearly see that the system is not bound. 

\begin{figure}[hbtp]
    \centering
    \includegraphics[width=0.95\columnwidth]{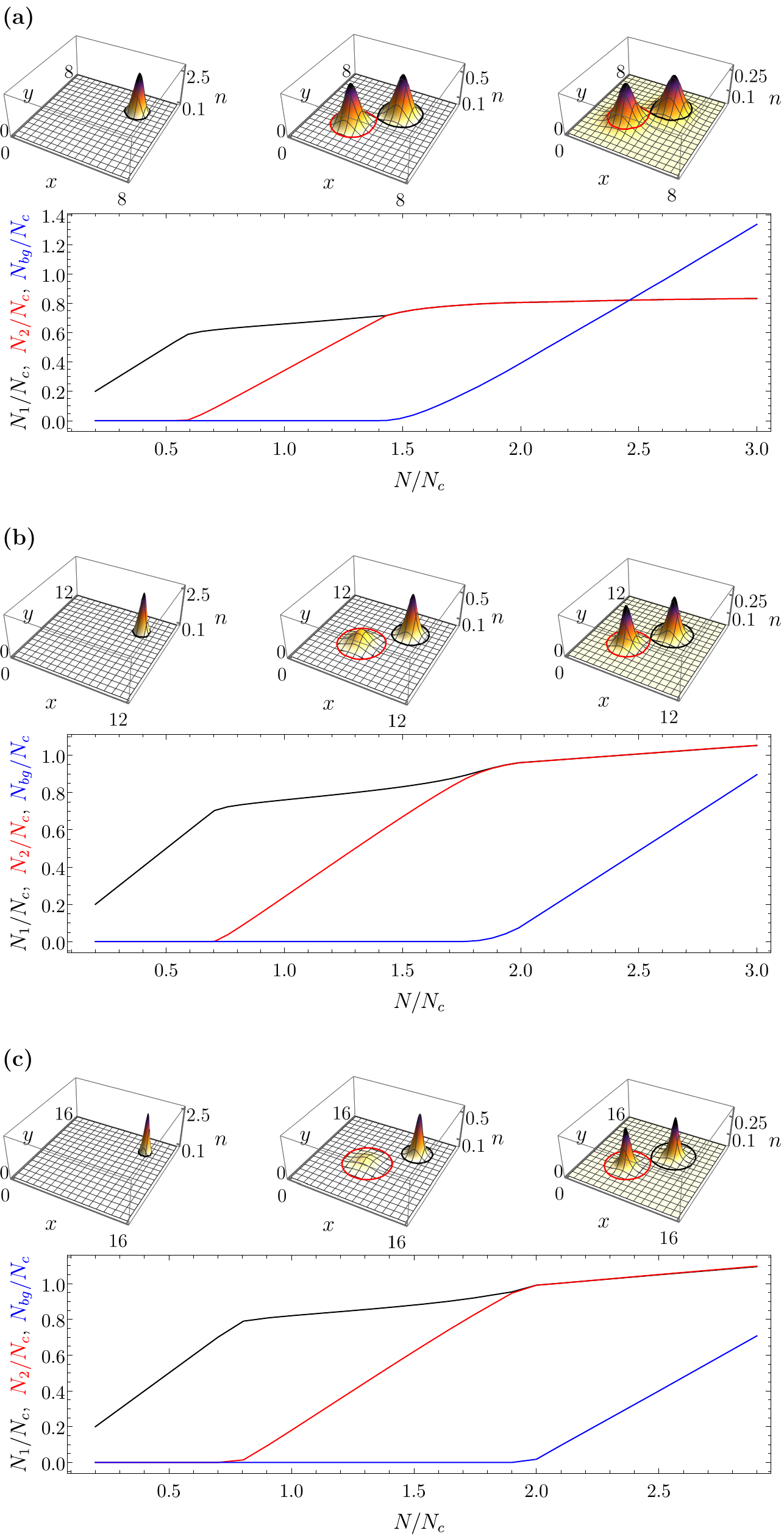}
    
    \caption{The heavy-fermion density profiles (3D plots) for the $N+2$ system confined in a square box of size $L\times L$, where $L$ is measured in units of the characteristic length scale $1/\sqrt{2m|\epsilon_1|}$. The vertical arrangement corresponds to $L=8$ (panel a), 12 (b), and 16 (c) and the horizontal arrangement corresponds to $N/N_c =$ 0.5 (left), 1.5 (middle), and 2.5 (right). The system always splits in individual $N+1$-type clusters which we encircle by the black and red contours in the 3D plots. The corresponding populations are shown in the 2D plots (black and red curves) as functions of $N$. The blue curves stand for the population of the free gas of heavy fermions (see text for precise definitions).
    }
    \label{fillingperpeak}
\end{figure}

Our numerical results are summarized in Fig.~\ref{fillingperpeak} where the panels (a), (b), and (c) correspond to the boxes with $L=8$, $L=12$, and $L=16$, respectively. $L$ is measured in units of $1/\sqrt{2m|\epsilon_1|}$. The 3D plots in each panel show the heavy-fermion density profiles $n({\bf r})$ for $N/N_c=0.5$ (left), $1.5$ (middle), and $2.5$ (right). The 2D plots in each panel show the populations of the first cluster $N_1$ (black), of the second cluster $N_2$ (red), and of the background gas of heavy atoms $N_{bg}$ (blue) in units of the saturation number $N_c$ for a $N+1$ cluster in free space. These numbers are defined as follows.

We identify the number of clusters by the number of local maxima of the field $n({\bf r})$. To quantify the population of a given cluster we find all points connected to the corresponding density peak where the heavy particle density $n$ is greater than a certain threshold. By default, the threshold is set to zero. If we obtain one or two disconnected clusters, their populations are defined as integrals of $n({\bf r})$ over the corresponding regions. This is what happens in all the leftmost 3D plots in Fig.~\ref{fillingperpeak} and in the middle 3D plots in panels (b) and (c). The black and red closed contours encircle the connected regions corresponding, respectively, to the first and the second cluster. If the clusters cannot be distinguished since the density is finite everywhere between the peaks, we raise the threshold to the saddle point value of $n({\bf r})$. In this manner we obtain two clusters with their edges touching at a single point (see the rightmost 3D plots in Fig.~\ref{fillingperpeak} and also in the middle 3D plot in Fig.~\ref{fillingperpeak}(a)). We again define $N_1$ and $N_2$ as integrals of $n({\bf r})$ over the corresponding regions and the rest is the background population $N_{bg}=N-N_1-N_2$.

With these definitions we can now identify three regimes of the system behavior according to how many of the quantities $N_1$, $N_2$ and $N_{bg}$ are not zero. The borders between these regimes depend on $L$, but the $L\rightarrow \infty$ scenario is rather clear. Namely, the first regime takes place for $N\lesssim N_c$ and corresponds to a single $N+1$ cluster plus a free light atom. We have checked that the light fermion is not bound to the cluster by looking at the delocalization of the corresponding orbital wave function as we increase the box size. Another consequence of the repulsion between the cluster and the free light atom is that the cluster is not in the middle of the box.  

In the second regime, for $N_c\lesssim N\lesssim 2N_c$, one cluster is saturated and we observe a gradual formation of the second one. The sizes and the peak heights of the two clusters are different. The size of the second cluster is expected to diverge at infinite $L$. This happens since our way of solving the problem implies that the coupling constant $g$ (or $\gamma$) makes the first saturated cluster stationary. This interaction is thus not sufficiently attractive to bind the second less populated cluster. For finite $L$ the size of the second cluster is determined by its interaction with the first cluster and with the walls of the box. We have checked that it increases with $L$.

For $N\gtrsim 2N_c$ we see two saturated clusters placed symmetrically on the box diagonal. The remaining heavy atoms are delocalized and take up all available space.

In the limit of infinite $L$ we expect $N_1=N$, $N_2=N_{bg}=0$ for $N<N_c$. Then, $N_1=N_c$, $N_2=N-N_c$, and $N_{bg}=0$ for $N_c<N<2N_c$. Finally, for $N>2N_c$ we have $N_1=N_2=N_c$ and $N_{bg}=N-2N_c$. Convergence to this limiting behavior with increasing the box size can be seen in the 2D plots in Fig.~\ref{fillingperpeak} as we go from (a) to (c). 

The relation of the energy to $\gamma_c$ given by Eq.~(\ref{Enminlead}) implies that the system is self bound and that it can minimize its energy by tuning its size. In this sense the problem in the box seems artificial because the box fixes the size and breaks the MF scaling invariance. Nevertheless, the point of doing this calculation is that {\it if the system formed a bound state, its density profile would eventually become insensitive to $L$}, at least for sufficiently large $L$. By contrast, we see that it is sensitive to the box size. In particular, we have checked that the distance between the two clusters, to a very good approximation, grows linearly with $L$. We cannot increase $L$ beyond $16$ as this is the limit of our numerical capabilities. To solve Eqs.~(\ref{SchrphiNN}) and (\ref{Dens}) we replace the continuum box by a lattice and the continuum Laplacian by its lattice version. What we show in Fig.~\ref{fillingperpeak} is the result of extrapolation to the continuum limit which, as we figure out, requires quite a fine grid. For example, the lattice that we use for $L=16$ contains up to $450\times 450$ grid points~\footnote{We do use the reflection symmetry with respect to the diagonal for our finest grids.}. We also have to run a few tens of thousands of iterations to reach the convergence in this case.

\section{Conclusion}
\label{Sec:Conc}

We obtain numerical evidence that two-dimensional clusters containing $N$ heavy fermions and two light fermions are not bound, at least, for sufficiently large $N$ and $M/m$. In this limit the MF theory becomes asymptotically exact and the system is controlled by a single dimensionless parameter $\alpha=4\pi N^2/(M/m)$ or, if we fix $M/m$, by the ratio $N/N_c$, where $N_c$ is the saturated number of heavy particles for the $N+1$ cluster. Another parameter which we have to introduce for unbound systems is the ratio of the box size to the length scale associated to the lowest orbital for the light atoms. In our numerical analysis we increase this ratio up to 16 observing that the system never stays localized, but expands in size with increasing the box dimensions.

Based on these calculations we conclude that the $N+2$ system is never bound and, if not confined, it always splits in $N+1$-type clusters and free atoms. In particular, for $N<N_c$, we are dealing with a $N+1$ cluster and a free light atom. For $N_c<N<2N_c$ this cluster is saturated and the second cluster starts to be formed. Finally, for $N>2N_c$ we have two saturated $N_c+1$ clusters plus a free gas of $N-2N_c$ heavy fermions.

This scenario is in contrast to what happens with the $N+2$ system in one dimension. For sufficiently small $\alpha_{\rm 1D}=(\pi^2/3)N^3/(M/m)$ there is a single $N+1$ cluster plus a free light, as we find also in two dimensions. However, for larger $\alpha_{\rm 1D}$ we deal with two identical clusters, which are not necessarily saturated and which, depending on $\alpha_{\rm 1D}$, may or may not be bound together.

Our statement about the absence of binding is based on the MF approximation valid in the limit of large $N$ and $M/m$. This is to say that we cannot exclude binding, for instance, in the $5+2$ system, but in this case one has to develop a method capable of addressing the fermionic seven-body problem in the regime where binding is either absent or very weak. It is also possible that $N+N_l$ clusters with $N_l>2$ may become bound in the MF regime. We leave these tasks for future studies.

\begin{acknowledgements}
We thank A. Tononi for useful discussions. We acknowledge support from ANR Grant Droplets No. ANR-19-CE30-0003-02. 
\end{acknowledgements}


\begin{thebibliography}{50}

\bibitem{KartavtsevMalykh} O. I. Kartavtsev and A. V. Malykh, \textit{Low-energy three-body dynamics in binary quantum gases}, J. Phys. B \textbf{40}, 1429 (2007). https://doi.org/10.1088/0953-4075/40/7/011

\bibitem{Blume} D. Blume, \textit{Universal Four-Body States in Heavy-Light Mixtures with a Positive Scattering Length}, Phys. Rev. Lett. \textbf{109}, 230404 (2012). https://doi.org/10.1103/PhysRevLett.109.230404

\bibitem{bazak2017} B. Bazak and D. S. Petrov, \textit{Five-Body Efimov Effect and Universal Pentamer in Fermionic Mixtures}, Phys. Rev. Lett. \textbf{118}, 083002 (2017). https://doi.org/10.1103/PhysRevLett.118.083002

\bibitem{Efimov} V. Efimov, \textit{Energy Levels of Three Resonantly Interacting Particles}, Nucl. Phys. A {\bf 210}, 157 (1973). https://doi.org/10.1016/0375-9474(73)90510-1

\bibitem{Castin} Y. Castin, C. Mora, and L. Pricoupenko, \textit{Four-Body Efimov Effect for Three Fermions and a Lighter Particle}, Phys. Rev. Lett. \textbf{105}, 223201 (2010). https://doi.org/10.1103/PhysRevLett.105.223201

\bibitem{PricoupenkoPedri} L. Pricoupenko and P. Pedri, \textit{Universal (1+2)-body bound states in planar atomic waveguides}, Phys. Rev. A \textbf{82}, 033625 (2010). https://doi.org/10.1103/PhysRevA.82.033625

\bibitem{liu2022} R. Liu, C. Peng, and X. Cui, \textit{Universal tetramer and pentamer in two-dimensional fermionic mixtures}, Phys. Rev. Lett. {\bf 129}, 073401. https://doi.org/10.1103/PhysRevLett.129.073401

\bibitem{Kartavtsev} O. I. Kartavtsev, A. V. Malykh, and S. A. Sofianos, \textit{Bound states and scattering lengths of three two-component particles with zero-range interactions under one-dimensional confinement}, J. Exp. Theor. Phys. \textbf{108}, 365 (2009). https://doi.org/10.1134/S1063776109030017

\bibitem{tononi2022} A. Tononi, J. Givois, and D. S. Petrov, \textit{Binding of heavy fermions by a single light atom in one dimension}, Phys. Rev. A \textbf{106}, L011302 (2022). https://doi.org/10.1103/PhysRevA.106.L011302

\bibitem{givois2023} J. Givois, A. Tononi, D.~S. Petrov, \textit{Self-binding of one-dimensional fermionic mixtures with zero-range interspecies attraction}, SciPost Phys. \textbf{14}, 091 (2023). https://doi.org/10.21468/SciPostPhys.14.5.091

\bibitem{givois2023_2} J. Givois, A. Tononi, D.~S. Petrov, \textit{Heavy-light N+1 clusters of two-dimensional fermions} arXiv:2310.11330 (2023). https://doi.org/10.48550/arXiv.2310.11330

\bibitem{Vlasov}
S.~N.~Vlasov, V.~A.~Petrishchev, and V.~I.~Talanov, \textit{Averaged description of wave beams in linear and nonlinear media}, Izv. Vyssh. Uchebn. Zaved. Radiofiz. {\bf 14}, 1353 (1971). https://doi.org/10.1007/BF01029467

\bibitem{Townes1964}R.~Y.~Chiao, E.~Garmire, and C.~H.~Townes, \textit{Self-trapping of optical beams}, Phys. Rev. Lett. {\bf 13}, 479 (1964). https://doi.org/10.1103/PhysRevLett.13.479



\end{thebibliography}
\end{document}